\documentclass[10pt]{iopart}
\usepackage{iopams}
\usepackage{amssymb}
\usepackage{graphicx}
\begin{document}

\letter{Saddle points in the chaotic analytic function and Ginibre characteristic polynomial}
    
\author{M R Dennis and J H Hannay}

\address{H H Wills Physics Laboratory, Tyndall Avenue, Bristol BS8 1TL, UK}

\begin{abstract}
Comparison is made between the distribution of saddle points in the 
chaotic analytic function and in the characteristic polynomials of 
the Ginibre ensemble. Realising the logarithmic derivative of these 
infinite polynomials as the electric field of a distribution of 
coulombic charges at the zeros, a simple mean-field electrostatic argument shows 
that the density of saddles minus zeros falls off as $\pi^{-1} |z|^{-4}$ from the 
origin. This behaviour is expected to be general for finite or 
infinite polynomials with zeros uniformly randomly distributed in the 
complex plane, and which repel quadratically.
\end{abstract}

It is well-known that there are several similarities between the 
distributions of zeros of the ensemble of random polynomials which 
tend to the chaotic analytic function (discussed by Hannay 1998, 1996, 
Bleher, Shiffman and Zelditch 2000, Forrester and Honner 1999, Leboeuf 1999, Bogomolny, Bohigas and 
Leboeuf 1996) and the eigenvalues of 
Ginibre matrices (Ginibre 1965, Mehta 1991 chapter 15), both in the finite and infinite cases. To 
be more precise, for $N$ large and possibly $\infty,$ we 
compare the zeros of the chaotic analytic function polynomials (caf 
polynomials)
\begin{equation}
    f_{N,{\rm caf}}(z) = \sum_{n=0}^N \frac{a_n z^n}{\sqrt{n!}},
    \label{eq:cafdef}
\end{equation}
where the $a_n$ are independent identically distributed complex circular 
gaussian random variables, with the eigenvalues of matrices in the 
Ginibre ensemble, defined to be $N\times N$ matrices with entries 
independent identically distributed complex circular gaussian 
random variables. The Ginibre analogue to equation (\ref{eq:cafdef}) is 
the characteristic polynomial $f_{N,{\rm Gin}}(z).$ The 
zeros of the two $f_N$ share the following properties, as discussed in 
the above references:
\begin{itemize}
    \item they are uniformly randomly distributed, with density 
    $\sigma = 1/\pi,$ within a disk centred on the origin 
    of the complex plane, which has radius $\sqrt{N},$ and  
    smoothed boundary (Ginibre with gaussian tail outside, caf with power law);
    \item within this disk, the distribution of zeros is statistically 
    invariant to translation and rotation;
    \item the statistical properties of the zeros at a fixed radius 
    $r$ do not change as $N$ increases, provided that  $\sqrt{N}\gtrsim r +\mathcal{O}(1);$
    \item the two-point correlation functions $g_{{\rm caf}}, 
    g_{{\rm Gin}}$ for zeros separated by distance $ |z_1 - z_2| = 
    r \lesssim \sqrt{N}-\mathcal{O}(1)$ 
    within the disk are given by 
    \begin{eqnarray}
	g_{{\rm caf}}(\sqrt{2} r) & = & [(\sinh^2 r^2 + r^4) \cosh r^2 - 2 r^2 \sinh 
	r^2]/\sinh^3 r^2, \label{eq:gcaf} \\
	g_{{\rm Gin}}(r) & = & 1 - \exp(-r^2). \label{eq:ggin}
    \end{eqnarray}
    Both of these functions exhibit quadratic repulsion at the origin, 
    are of order 1 (uncorrelated) for $r \gtrsim 1,$ and satisfy a 
    screening relation: the integral of $\sigma g$ over the plane, after subtracting 
    the uniform background density, is $-1;$
    \item the distribution of Ginibre zeros is equivalent to a two 
    dimensional $N$-charge Coulomb gas in a harmonic oscillator potential 
    $|z|^2/2,$ at a temperature corresponding to $\beta = 2,$ whereas the 
    caf zeros have an additional $N$-body potential.
\end{itemize}
Despite these similarities, there is no obvious explicit relation 
between these two ensembles.

\begin{figure}
\begin{center}
\includegraphics*[width=12cm]{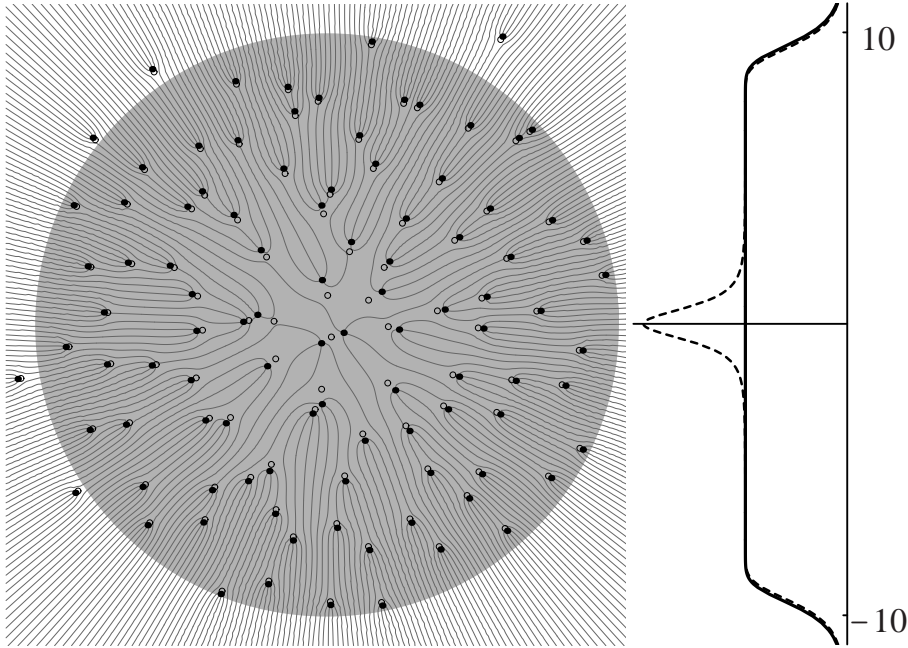}
\end{center}
    \caption{A sample distribution of zeros and saddles for a caf 
    polynomial with $N = 100.$ The zeros are represented by filled 
    circles, the saddles by empty circles. The large disk has 
    radius 10, indicating the area in which the distribution is 
    isotropic. Lines of constant argument (equally spaced by $\pi/2$) are given by the curves; in the electrostatic analogy, they are electric field lines. The plot to the right shows the theoretical density of zeros $\sigma$ (full line) and saddles $\rho$ (dashed line) for $N = 100,$ showing the saddle bump at the origin and the disk boundary smoothing (the saddle boundary lies just inside the zero boundary, giving one {\em fewer} saddles than zeros in all.}
    \label{fig:disk}
\end{figure}

In this work, we consider the distribution of the saddle points of the polynomials $f_{{\rm caf}}, f_{{\rm Gin}}$ (dropping the 
suffix $N$ unless necessary), that is, zeros of the derivative $\rmd f/\rmd z.$ The behaviour of the caf saddles is easy to determine, the Ginibre saddles less so. 
Numerical experiment shows that in both cases the saddle distribution roughly mimics the zero distribution, except for a surplus near the origin and a deficit 
near the disk edge. 
Using the electrostatic analogy, we shall see that the density of the saddles 
minus zeros has the same $1/\pi |z|^4$ tail away from the 
origin for each case, due to the quadratic repulsion of the zeros. 
Also, the cause and extent of the edge deficit will be clear.

The density of caf saddles $\rho$ can be found by replacing $f_{\rm caf}$ with $f_{\rm 
caf}'$ in the formula for the density of zeros  (Hannay 1996, Nonnenmacher and Voros 1998 equation (69)). When $|z|\lesssim \sqrt{N} -\mathcal{O}(1),$ the $N \to \infty$ formulae may be applied. In this case, the zero density $\sigma$ is $1/\pi$ (as mentioned above), and the saddle density is 
\begin{equation}
    \rho = \pi^{-1} \partial_{z} \partial_{z^*} 
    \langle f_{{\rm caf}}'(z) f_{{\rm caf}}'(z)^* \rangle = 
    \pi^{-1} (1 + (1+|z|^2)^{-2}).
    \label{eq:saddens}
\end{equation}
This has, in addition to a uniform $1/\pi$ density, a `bump' 
in the vicinity of the origin, which integrates to 1, and the saddle 
surplus is $\rho - \sigma = (\pi(1+|z|^2)^2)^{-1},$ 
which decays like $\pi^{-1} |z|^{-4}.$ For finite $N,$ 
of course, an order $N$ polynomial has $N-1$ saddles, not $N+1,$ implying that $\rho \approx 0$ for $|z|^2 \gtrsim N-2,$ which is indeed the case 
when the calculation in equation (\ref{eq:saddens}) is performed for 
finite $N.$ The pattern of zeros and saddles for a sample caf 
polynomial with $N = 100$ is shown in figure \ref{fig:disk} (the 
pattern of Ginibre zeros and saddles look very similar), along with the corresponding plots of $\sigma$ and $\rho;$ the most obvious feature of this 
distribution is that zeros and saddles tend to occur in `bonded' pairs, with 
the saddle positioned radially inwards from the zero, and the 
bond length decreases as $|z|$ increases. The pairing 
breaks down near the origin, indeed the `extra' unbonded zero is in this neighbourhood.

\begin{figure}
\begin{center}
\includegraphics*[width=12cm]{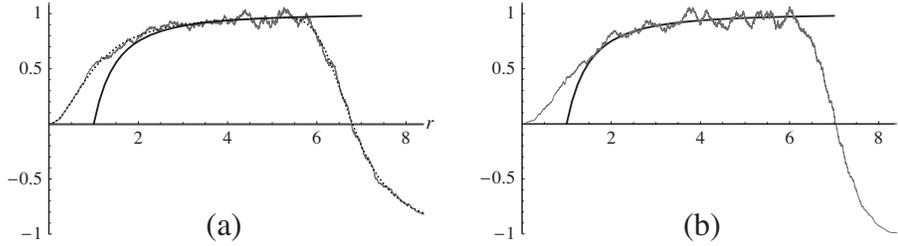}
\end{center}
    \caption{Plots of $\mathcal{N}(r)$ compared against the asymptotic
    theoretical fit: (a) caf, fitted with $1-1/r^2$ as in 
    (\ref{eq:neqapp}) (solid line) and the exact $\mathcal{N}$ from 
    (\ref{eq:saddens}), (\ref{eq:neq}) (dashed line); (b) Ginibre, fitted with 
    $1-1/r^2$ as in (a). The data in each case are for 
    1000 random samples with $N = 50$ (numerical errors tend to 
    appear for Ginibre with $N$ significantly larger).}
    \label{fig:nnum}
\end{figure}

It proves to be more convenient to work with the cumulative 
number of saddles minus zeros within a radius $r = |z|,$ denoted by 
$\mathcal{N}(r),$ which is given by
\begin{eqnarray}
    \mathcal{N}(r) & =  & 2 \pi \int_0^r (\rho - 
    \sigma ) r' \, \rmd r'\label{eq:neq} \\
    & \approx & 1 - 1/r^2 \quad \hbox{for caf, with $1\ll r \lesssim 
    \sqrt{N}-\mathcal{O}(1).$} 
    \label{eq:neqapp}
\end{eqnarray}
It is our aim to show, using mean-field electrostatics, that this expression 
holds generally for polynomials with zeros distributed according to 
the bullet points above, including the Ginibre polynomials 
$f_{{\rm Gin}}.$ Figure \ref{fig:nnum} shows numerical plots of 
$\mathcal{N}(r)$ for simulations of both the caf and Ginibre 
polynomials, the caf fitting (\ref{eq:neqapp}) as well as the exact theoretical curve 
(putting the finite form of (\ref{eq:saddens}) into (\ref{eq:neq})), the Ginibre, 
fitting with (\ref{eq:neqapp}).

We start by recalling that the (two dimensional) electric field is 
represented by the logarithmic derivative of 
the polynomial $f,$ with $N$ zeros $z_j,$ which electrostatically 
represent $N$ unit charges at positions $z_j$ in the plane. The logarithmic 
derivative $\mathcal{E}$ of $f$ is
\begin{equation}
    \mathcal{E}(z) = \frac{\rmd \log f(z)}{\rmd z} = \sum_{j=1}^N \frac{1}{z-z_j},
    \label{eq:logdere}
\end{equation}
which is Coulomb's law in two dimensions (setting $q/2 \pi 
\varepsilon_0 = 1$), and holds when there are arbitrarily many 
zeros/charges. The direction of the electric vector 
$\bi{E}$ is given by $\mathcal{E}^*,$ i.e. $\bi{E} = 
({\rm Re}\mathcal{E},-{\rm Im}\mathcal{E}),$ and the direction of the field lines is 
given by the contours of constant argument of $f.$ The magnitude 
of the field $|\mathcal{E} | = |\bi{E}|,$ and saddles are the places 
where this is zero. It will also be useful to define the field due to 
all charges excluding a particular one at $z_i,$
\begin{equation}
    \mathcal{E}_i(z) = \sum_{j \neq i, j=1}^N \frac{1}{z-z_j}.
    \label{eq:ei}
\end{equation}

Saddles may therefore be thought of as the places where the field from a zero/charge balances the background field from all of the others. 
Consider a particular zero at $z_i$ with $\sqrt{N}-\mathcal{O}(1) \gtrsim |z_i| \gg 1.$ 
By Gauss's law, the statistically disklike distribution of the zeros implies that the average field at $z_i$ due to the other charges is that due to a charge $\sigma \pi |z_i|^2$ at the origin (only zeros with modulus less than $|z_i|$ are relevant). 
This justifies  electrostatically the observation from figure \ref{fig:disk} that, for sufficiently large $z_i,$ zeros and saddles are almost always paired, and that the saddle $s_i$ paired with the zero $z_i$ is near $z_i,$ on the straight line between the zero and the origin. 
We denote the real positive `bond length' $|z_i - s_i|$ by $b(s_i) = b_i.$
This bond length function will be calculated below on the basis of the electrostatic model, ignoring statistical fluctuations (implying that $b(s_i)$ is radial). 
This will yield $\mathcal{N}(r)$ by finding the number $\#(r)$ of bonds crossed by a circle of radius $r.$ 
Since for a polynomial, $\mathcal{N}(\infty) = -1$ and $\#(\infty)=0,$ we can set $\mathcal{N}(r) = \#(r)-1.$

The number $\#(r)$ is simply the number of zeros in the annulus whose 
inner radius $r$ is $|s_i|,$ and whose outer radius is $|s_i| + b(s_i);$ 
any zero $z_j$ in this area will have a saddle with $|s_j|<|s_i|,$ 
and its bond $b_j$ will cross the circle of radius $|s_i|.$ 
Therefore, with $|s_i| = r,$ and realising that the zero density $\sigma$ is uniform,
\begin{eqnarray}
    \mathcal{N}(r) +1 & = & \#(r) = \sigma \pi [(b(r) + r)^2 - r^2] 
    \nonumber \\
    & = & b(r)^2 + 2 r b(r). \label{eq:bondno}
\end{eqnarray}

The problem remains of how to calculate $b(r).$ The crudest 
approximation is to balance 
radially the field from $z_i$ with the field from the rest of the charges 
$\mathcal{E}_i,$ again ignoring fluctuations. Using Coulomb's 
law, and applying Gauss's law at $|z_i|,$ gives 
$1/b_i = |z_i|$ or $b_i = 1/|z_i|.$ Solving $b_i$ in terms of 
$|s_i|$ and putting into equation 
(\ref{eq:bondno}) does indeed give the required leading order terms 
$1- 1/r^2$ in (\ref{eq:neqapp}), but we have made two approximations in 
this argument which affect the value of $b_i$ at the required 
order. We show below that these two effects cancel each other.

The first approximation made is that Gauss's law should really be applied at the saddle 
radius $|s_i|,$ not the zero radius $(|s_i|+ b(s_i)).$ Substituting 
this value into (\ref{eq:bondno}), however, gives 
$\mathcal{N}(r) = 1 + 1/r^2,$ not $1 - 1/r^2$ as desired. The second 
approximation is that the repulsion from $z_i,$ embodied by the correlation function 
$g,$ has been neglected.

The crude approximation assumed that the field due to 
the charges other than $z_i$ to be due to a uniform jellium of density 
$\sigma.$ However, we know from the two-point correlation function 
$g$ (with the properties above) that zeros are repelled quadratically from a given one, and 
the background jellium is `dented' around $z_i,$ with the shape of 
the dent given around $z_i$ by $g(|z-z_i|).$ The correct field to 
use, in this case, is the mean of (\ref{eq:ei}), not (\ref{eq:logdere}). 
Gauss's law, now applied 
to the dent (since it is circularly symmetric around $z_i$), effectively 
weakens the field from $z_i.$ Including this correction as well, we have 
the implicit expression for $b$ (omitting $i$ subscripts):
\begin{equation}
    \frac{1}{b} -\frac{2 \pi \sigma}{b}\int_0^b (1-g(b')) b' \, \rmd b' = s.
    \label{eq:bimp}
\end{equation}
Since $b$ is very small (of the order of $1/s$), $g(b)$ is 
proportional to $b^2$ due to repulsion, and the $g(b')$ part of the integrand 
may be neglected, integrating to $b^2/2$ to leading order; thus (\ref{eq:bimp}) 
implies  
\begin{equation}
    b_i \approx \frac{1}{|s_i|+b_i} = \frac{1}{|z_i|},
    \label{eq:bexp}
\end{equation}
which is the same as that crudely derived above. Using this mean-field jellium 
approximation, we have therefore justified the numerical observation 
that $\mathcal{N}(r) \approx 1 - 1/r^2,$ (equation (\ref{eq:neqapp})) and therefore the density of 
saddles minus zeros, to leading order, decays as $1/\pi r^4.$ We make the following 
observations:
\begin{itemize}
    \item The main objection to this derivation, of course, is that 
    statistical fluctuations have been neglected throughout the 
    discussion. By Cauchy's theorem, the exact saddle excess 
    $\mathcal{N}(r)$ is $2\pi r \langle \sum (r-z_j)^{-2}/\sum 
    (r-z_j)^{-1}\rangle,$ which we do not know how to evaluate. Both 
    numerator and denominator fluctuate violently, and approximating 
    the average of the ratio by the ratio of the averages fails; it 
    appears that fluctuations deny us information of the density beyond 
    the $r^{-4}$ leading order term obtained. Incidentally, for less 
    violently fluctuating fields, for example from charges on a 
    unit circle instead of a disk (appropriate for CUE (Mezzadri, 2002)), 
    this approximation succeeds in reproducing the leading behaviour 
    $\mathcal{N}(r) \approx 2 \pi r \langle \sum (r-z_j)^{-2} \sum 
    (r-z_j)^{-1*}\rangle/\langle|\sum (r-z_j)^{-1}|^2\rangle = 
    1/(1-r)$  for $1/N \ll (1-r) \ll 1$ (the ratio has been 
    rationalised since the denominator is otherwise zero by Gauss's 
    law).
    \item Fluctuations aside, our result follows only by assuming 
    repulsion between the zeros, and it is easy to check numerically 
    that for polynomials with zeros distributed 
    completely at random in a disk (Poisson distribution),  
    $\mathcal{N}(r)$ does not have the form (\ref{eq:neqapp}). The mathematical form of $\mathcal{N}(r)$ is 
    not known for the Poisson distribution.
    \item We remark that in the infinite caf case, the density of saddles 
    minus zeros, using equation (\ref{eq:saddens}), is uniform on the 
    sphere upon stereographic projection (Needham, 1997, p 146); 
    the distribution of neither zeros nor saddles is separately uniform. 
\end{itemize}

\ack We are grateful to F. Mezzadri for discussions, and to S.D. 
Maplesden for performing preliminary calculations as part of a 
final year undergraduate project. MRD is supported by the Leverhulme 
Trust.

\References

\item[]
Bleher P, Shiffman B and Zelditch S 2000
\newblock {Universality and scaling between zeros on complex manifolds}
\newblock {\em Invent.math.} {\bf 142} 351--95

\item[]
Bogomolny E, Bohigas O and Leboeuf P 1996
\newblock {Quantum chaotic dynamics and random polynomials}
\newblock {\em J.Stat.Phys.} {\bf 85} 639--79

\item[]
Forrester P J and Honner G 1999
\newblock {Exact statistical properties of complex random polynomials}
\newblock {\em J.Phys.A:Math.Gen.} {\bf 32} 2961--81

\item[]
Ginibre J 1965
\newblock {Statistical ensembles of complex, quaternion and real matrices}
\newblock {\em J.Math.Phys.} {\bf 6} 440--9

\item[]
Hannay J H 1996
\newblock {Chaotic analytic zero points: exact statistics for a random spin
  state}
\newblock {\em J.Phys.A:Math.Gen.} {\bf 29} L101--5

\item[]
\dash 1998
\newblock {The chaotic analytic function}
\newblock {\em J.Phys.A:Math.Gen.} {\bf 31} L755--61

\item[]
Leboeuf P 1999
\newblock {Random analytic chaotic eigenstates}
\newblock {\em J.Stat.Phys.} {\bf 95} 651--64

\item[]
Mehta M L 1991 
\newblock {\em Random Matrices}
\newblock (Academic Press)

\item[]
Mezzadri F 2002
\newblock Random matrix theory and the zeros of $\zeta'(s),$
\newblock {\em J.Phys.A:Math.Gen.} in press

\item[]
Needham T 1997 
\newblock {\em Visual Complex Analysis}
\newblock (Oxford University Press)

\item[]
Nonnenmacher S and Voros A 1998
\newblock {Chaotic eigenfunctions in phase space}
\newblock {\em J.Stat.Phys.} {\bf 92} 431--518

\endrefs

\end{document}